\documentclass[pra,a4paper,twocolumn,showpacs,superscriptaddress,amsmath,amssymb,floatfix,amstex]{revtex4}
\newcommand{\ket}[1]{| #1\rangle}        		%
\newcommand{\bra}[1]{\langle #1}         		%
\newcommand{\brac}[1]{\langle #1|}         		%
\newcommand{\pd}{\partial}                  		%
\newcommand{\nx}[1]{\mathbf{#1}}	    		%
\newcommand{\su}[1]{\text{\textbf{\textsf{#1}}}}	%
\newcommand{\s}{\su{S}}					%
\newfont{\Bb}{msbm10}					%
\newcommand{\II}{I}				%
\newcommand{\tr}{{\rm Tr}}				%
\newcommand{\equa}[1]{Eq.~(\ref{#1})}       		%
\newcommand{\fig}[1]{FIG.~\ref{#1}}         		%
\newcommand{\sect}[1]{Sec.~\ref{#1}}			%
\usepackage{graphicx,graphics,wrapfig,rotating}	
\usepackage{dcolumn}				
\usepackage{bm,pstricks}					
\usepackage{hyperref}

\usepackage{mathptmx,helvet}	
\newrgbcolor{darkblue}{0.165 0.1 0.43}
\newrgbcolor{lcolor}{0. 0. 0.80}
\newrgbcolor{gcolor2}{.80 .80 1.}
\newrgbcolor{terra}{0.73 0.28 0.22}
\newrgbcolor{qcggreen}{0.71 0.78 0.68}
\newrgbcolor{qcgyellow}{1. 0.85 0.5}
\newrgbcolor{darkgreen}{0.25 0.46 0.22}
\begin{document}


\title{Phase space contraction and quantum operations}

\author{Ignacio Garc\'{\i}a-Mata}
\email{garciama@tandar.cnea.gov.ar}
\affiliation{%
Departamento de F\'\i sica, Comisi\'on Nacional de Energ\'\i a At\'omica.
Av del Libertador 8250 (1429), Buenos Aires, Argentina
}
\author{Marcos Saraceno}
\affiliation{%
Departamento de F\'\i sica, Comisi\'on Nacional de Energ\'\i a At\'omica.
Av del Libertador 8250 (1429), Buenos Aires, Argentina
}%
\affiliation{%
Escuela de Ciencia y Tecnolog\'\i a, Universidad Nacional de San
Mart\'\i n. Alem 3901 (B1653HIM), Villa Ballester, 
Argentina}
\author{Mar\'{\i}a Elena Spina}
\affiliation{%
Departamento de F\'\i sica, Comisi\'on Nacional de Energ\'\i a At\'omica.
Av del Libertador 8250 (1429), Buenos Aires, Argentina
}
\author{Gabriel Carlo}
\affiliation{%
Center for Nonlinear and Complex Systems, Universit\`a degli Studi dell'Insubria and Instituto Nazionale
per la Fisica della Materia, Unit\`a di Como, Via Valleggio 11, 22100 Como, Italy
}
\date{\today}
\begin{abstract}
We give a criterion to differentiate between dissipative and diffusive quantum operations. It 
is based on the classical idea that dissipative processes contract volumes in phase space.
We define a quantity that can be regarded as ``quantum phase space contraction rate'' and which is related to a fundamental
property of quantum channels: non-unitality. We relate it to other properties of the channel and also show a simple example of
dissipative noise composed with a chaotic map. The emergence of attaractor-like structures is displayed.
\end{abstract}
\maketitle
\section{Introduction}
Quantum dissipative processes  in the context of quantum optics  \cite{gardiner}, 
of superradiance \cite{haakebook}, of cooling mechanisms in ion traps \cite{iontraps} and open nanostructures
\cite{mahler}   
have been widely studied using a master equation approach. The corresponding master 
equation for each case is derived by modeling a microscopic interaction between the 
system (according to the case: an oscillator, a large spin, etc.) and a heat bath 
representing the environment.
Under standard approximations (e.g. Born-Markov, random wave)\cite{gardiner} 
one ends with a Lindblad master equation, 
which automatically insures the preservation of positivity \cite{lind,gks}. 
For general dissipative processes the operators in the master equation are not normal.
 
In this article dissipation is understood in analogy with classical dynamics, that is, we call 
dissipative any process in which phase space volume is not preserved.
We have to keep in mind that this definition is different from the concept of 
`dissipativity' which is often found in the literature.  For example, 
in \cite{lind} Lindblad calls dissipative maps the generators of completely positive semigroups, 
that is, dissipativity is linked to irreversibility. For Lidar et al.\cite{lidar} the condition of 
dissipativity is related to strictly purity decreasing, and shows its close relation 
to unitality. The concept of phase space contraction is well defined in a classical context as the
divergence of the velocity field. In quantum mechanics this divergence can be related to the sum of the commutators of the Lindblad
generators in the classical limit. \cite{percival}

The evolution of the density matrix of the system between two given times, that is,
the transformation that takes $\rho(t)$ to $\rho (t + \delta t) $ can be 
obtained by integration of the master equation.  This leads to a superoperator which 
can be written in an operator sum or Kraus form \cite{kraus}. The detailed procedure depends on 
each particular process and can be mathematically complicated.
In this work we follow an alternative approach  to the description of dissipative 
quantum noise which consists in directly modeling the superoperator in its Kraus 
representation. The formalism of quantum operations to describe open systems, 
which is especially well adapted in the context of quantum information and quantum 
computing \cite{chuang,preskill},  has also been used for open quantum maps 
\cite{bian,garma,garma2,colo,sloppy,nonn}, which are quantum maps in the usual sense with some
non-unitary noise that can be considered the effect of an interaction with some environment.

In \cite{colo} 
the authors analyze several models for purely diffusive noises and show that  
generalization of simple channels known in quantum information theory (depolarizing 
and phase damping channels)
can be written as an incoherent sum of translations in phase space. Since the 
translation operators are unitary the preservation of the trace implies that 
the noise superoperator is unital (i.e. identity preserving). In this paper we 
focus on non-unital channels in a finite dimensional Hilbert space
and show that their action can be interpreted as a contraction of the phase space
volume leading to dissipative dynamics.

This paper is organized as follows. In \sect{sec:opensys} we briefly go through the formalism of 
the Lindblad master equation and the superoperator approach.
A parameter measuring phase space contraction in the quantum operations formalism 
is introduced in \sect{sec:contraction}. Some examples of quantum diffusive and dissipative processes are  
given in\sect{sec:examples}.
In \sect{sec:SDC} we give a simple dissipative channel as an example and show the effects when
composed with a unitary map in \sect{sec:unitary}.
\section{Master equation and Quantum dynamical semigroups}
	\label{sec:opensys}
Open systems involve non-unitary evolution of states into probabilistic
ensembles of states, or mixed states. The description of mixed states is given 
by the density operator $\rho$. 
Usually the dynamics of open quantum systems is described by a Hamiltonian 
\begin{equation}
H=H_s+H_r+H_{i}
\end{equation}
consisting of the dynamics of the system ($s$) and the reservoir ($r$) and an 
interaction term.
If the Markov approximation holds for the reservoir, then the problem 
can be reduced
to a master equation for the density matrix of the {\em system\/}, 
\begin{equation}
	\label{eq:master0}
\frac{\pd{\rho}}{\pd t}={\cal L}(\rho)
\end{equation}
which in general can be written (in the interaction picture)
in GKS form \cite{gks} and in turn can be simplified into the 
so-called Lindblad \cite{lind} form
\begin{equation}
	\label{eq:master}
\frac{\pd{\rho}}{\pd t}=\frac{1}{2\hbar}
\sum_i \left\{[L_i,\rho L_i^\dag]+
[L_i\rho,L_i^\dag]\right\}
\end{equation}
where the Lindblad operators $L_i$ are system operators which are in principle 
arbitrary and which characterize the noise.


The formal solution of the master equation is
\begin{equation}
	\label{eq:sol}
\rho(\Delta t)=e^{\Delta t{\cal L}}\rho(0)
\end{equation}
so ${\cal L}$ is the generator
of a one-parameter family of completely positive (CP), trace 
preserving (TP) linear maps $\su{S}$ called 
{\em quantum dynamical semigroup\/}\cite{lind,gks}. These operators 
acting on the space of density matrices (the subspace of positive semidefinite operators 
include the density matrices) are given such names as superoperators, 
quantum operations or channels depending on the context 
in which they appear. We use any of those names indistinctively. 
Moreover in this work we
consider only discrete time CP maps which can (but not necessarily do) arise from an integration of a Lindblad
equation through time $\Delta t$.

It is well known \cite{chuang,preskill} that a completely positive superoperator can 
be written in the {\em operator sum\/} or
Kraus \cite{kraus} form. That is, there exists a set of operators $M_\mu$ such that 
\begin{equation}
	 \label{eq:kraus}
\su{S}(\rho)=\sum_\mu M_\mu\rho M_\mu^\dag,
\end{equation}
where $\mu\le N^2$, with $N$ the dimension of the state space. the map $\su{S}$ is TP
if
\begin{equation}
	\label{eq:tp}
\sum_\mu M_\mu^\dag M_\mu=1.
\end{equation}

In the analysis that follows pure states are vectors in finite dimensional Hilbert space ${\cal H}_N$  (of dimension
$N$) and therefore density operators are represented as complex $N\times N$ matrices.
A finite Hilbert space of dimension $N$ is naturally associated to a classical phase space of finite volume
(that we normalize to unity). The dimension is then given by the semiclassical rule $N=1/h$
The volume occupied by a state $\rho$ is well represented by the purity
$\tr[\rho^2]$. 
The simplest examples are the sphere (for angular momentum applications) and the torus (for quantum
maps). The association of quantum density matrices to distributions in phase space can be done in several
well known ways via the Weyl-Wigner, Husimi or Kirkwood representations.
\section{Quantum phase space contraction}
	\label{sec:contraction}
The standard definition of dissipation in classical mechanics is related 
to the contraction of phase space volumes given by the divergence of
 the drift vector entering the Fokker Plank equation \cite{licht}. 
 
At the level of the quantum master equation it is well understood 
how the properties of the Lindblad operators determine whether
 dissipation  will be present in the corresponding classical dynamics 
\cite{brodier,percival}. 
In \cite{strunz}  Strunz and Percival derive a Fokker Planck equation  by taking
the semiclassical limit of a quantum master equation of the type (\ref{eq:master})
and show how non-Hermitian Lindblad operators 
lead to classical dissipation (and quantum fluctuations),
 while Hermitian ones describe diffusion (these could also 
have a contribution to dissipation, but in usual physical 
applications they do not).  They obtain an expression to lowest order in $\hbar$ of 
the phase space divergence of the vector drift 
$\mathbf{A}$ in terms of the Poisson
 brackets of the Wigner-Moyal transforms $L_i(q,p)$ of 
the Lindblad operators 
\begin{equation}
	\label{eq:divA}
{\rm div}\,\mathbf{A}=-i\sum_i\{L_i(q,p),L_i^\dag(q,p)\}.
\end{equation}
Given the correspondence between Poisson brackets and commutators
it is clear from this equation that dissipative processes (${\rm div}\, \mathbf{A}\neq 0$) 
correspond to non normal Lindblad operators (i.e. $L_i L_i^\dag=L_i^\dag L_i$) 
in the master equation (\ref{eq:master}), that is to {\em non-unital\/} 
processes which by definition satisfy ${\cal L}(I)\neq 0$ 
or, at the level of superoperators $\s(I)\neq I$.

Dissipation is then described by non-unital quantum operations.
On this basis we  now introduce a dissipation parameter, which measures
 the non unitality of a given quantum channel,
\begin{equation}
	\label{eq:eta}
\eta=\frac{\tr\left[\left(\su{S}(\rho_I)-\rho_I\right)^2\right]}{\tr[(\rho_I)^2]}=
N\,\tr[\Gamma^2],
\end{equation}
where $\rho_I=I/N$ is the maximally mixed state and
\begin{equation}
	\label{eq:defA}
\Gamma=\s(\rho_I)-\rho_I=\frac{1}{N}\sum_\mu[M_\mu,M_\mu^\dag]
\end{equation}
is a traceless, Hermitian operator which gives a measure of the non-normality of the Kraus
operators $M_\mu$. 

Dissipation leads to a concentration of probability in some states (with a
consequent reduction in others). In this sense, a non-vanishing $\eta$
implies that a {\em contraction\/} of the
available phase space volume with respect to the uniform distribution 
has been achieved. It is this contraction that is measured by the parameter $\eta$.

This operator has significance in a phase space representation. For example, its Husimi function
\begin{equation}
\Gamma(z,z^*)=\frac{\brac{z}\Gamma\ket{z}}{\bra{z}\ket{z}},
\end{equation}
where $\ket{z}$ is the coherent state centered at $z=(q,p)$ (and $z^*$ implies complex
conjugate), gives a local description of the
dissipation process. However, this is not the only possibility as other
operator bases may be used to provide different pictures. For example, for quantum information applications the
Pauli \cite{chuang} basis plays a prominent role\footnote{For the usual one-qubit amplitude damping channel
\cite{chuang}, written in terms of the Pauli matrices
$\{\sigma_x,\sigma_y,\sigma_z\}$ and the identity $I$, the Kraus operators are
$M_0=((1+\sqrt{\smash[b]{1-\gamma}})I+(1-\sqrt{\smash[b]{1-\gamma}})\sigma_z)/2$ and
$M_1=\sqrt{\smash[b]{\gamma}}\,(\sigma_x+i\sigma_y)/2$. In this basis we have
$\Gamma=\gamma \sigma_z$ and $\eta = \gamma^2$.}. 

A more precise connection between the parameter $\eta$ and a classical
contraction rate can be made by writing \equa{eq:eta} in terms of the variation
of the purity
\begin{equation}
p_n(\rho)=\tr[\rho_n^2]=\tr[(\su{S}^{n}(\rho))^2],
\end{equation}
 at ``time'' $n$. Expanding the square in \equa{eq:eta} it follows that
\begin{equation}
\eta=\frac{\tr\left[\s(\rho_I)^2-(\rho_I)^2\right]}{\tr[(\rho_I)^2]}=
\frac{p_1(\rho_I)- p_0(\rho_I)}{p_0(\rho_I)}.
\end{equation}

This quantity can also be related to the Lindblad operators by taking $\Delta t$ in \equa{eq:sol}
small enough so that a quadratic expansion of the purity
\begin{equation}
p_t(\rho_I)=p_0(\rho_I)+\dot{p}_0(\rho_I) \Delta t+\frac{1}{2}\ddot{p}_0(\rho_I)\Delta t^2+\cdots
\end{equation}
is valid.
Since the first derivative of the purity for $\rho_I$ vanishes, 
we find (for $\Delta t=1$) 
\begin{equation}
\frac{p_1(\rho_I)-p_0(\rho_I)}{p_0(\rho_I)}=\frac{1}{\hbar^2}\sum_{ij}\tr\left([L_i,L_i^\dag][L_j,L_j^\dag]\right)
\end{equation}
In  the classical limit, from Eqs. (\ref{eq:divA}) and (\ref{eq:eta}) we get:
\begin{equation}
\eta\xrightarrow[\hbar\to 0]{} \int({\rm div}\,\mathbf{A})^2 dqdp
\end{equation}
and thus $ \eta $ can be understood as a global
contraction rate. 

There is another way to interpret the parameter $\eta$.
The map $\s$ acts linearly on the space of operators. Therefore it has  associated a matrix representation.
If we consider an orthonormal basis of $N^2$ operators $\Lambda_i$,
such that $\Lambda_0=\II/\sqrt{N}$ and the rest are traceless operators, the matrix representation of $\s$
takes the form
\begin{equation}
\label{eq:affine}
[\s]=
\left[
\begin{array}{c|ccc}
1& &\nx{v}_2& \\ \hline
 & & & \\
\nx{v}_1 & &\nx{M}&
\end{array}
\right]
\end{equation}
where
\begin{equation}
[\s]_{ij}=\tr[\Lambda_i^\dag\s(\Lambda_j)].
\end{equation}
The matrix $\nx{M}$ is  $(N^2-1)\times(N^2-1)$ and $\nx{v}_1$ and $\nx{v}_2$ are column and row vectors
respectively (we use the square brackets to identify the representation of a superoperator or an operator in
this basis). This is the so-called affine representation.
The trace
preserving condition is met if $\nx{v}_2=0$ while $\nx{v}_1=0$ implies that the map is unital. 
So if $\nx{v}_1\neq 0$ in \equa{eq:affine}, the contraction parameter is
\begin{equation}
\eta=\frac{([\s][\rho_I]-[\rho_I])^\dag ([\s][\rho_I]-[\rho_I])}{[\rho_I]^\dag[\rho_I]}.
\end{equation}
Now in this basis 
\begin{equation}
[\rho_I]=\left(
\begin{array}{c}
1/\sqrt{N}\\
 0\\
 \vdots\\
 0\\
\end{array}
\right)
\end{equation}
which leads to
\begin{equation}
\eta=\nx{v}_1^\dag\nx{v}_1
\end{equation}
and the contraction information is contained in the first column of $[\s]$.
In a way $\eta$ measures the non-symmetric form the matrix takes in the affine representation. 

It is worth
remarking that while the left invariant eigenstate of $[\s]$ is $[\rho_I]$ the right invariant eigenstate is
something different which in the case of the noise composed with a unitary, possibly chaotic, map can take
a very involved form resembling the fractal structure of a strange attractor. An example of this is given
in \sect{sec:unitary}. 
\section{Examples}
	\label{sec:examples}
In this section we analyze some examples of purely diffusive and dissipative 
noise models found in the literature to illustrate 
the classification introduced above and compute the corresponding contraction parameter $\eta$.

Our definition does not make any explicit reference to the classical phase space, it only refers to finite dimensional
Hilbert space. In what follows we choose to use the torus which can be represented as a square of unit
side, with periodic boundary conditions.
Therefore the superoperator $\su{S}$ acts on the space of complex
$N\times N$ matrices.
\subsection{Random unitary processes}
Trace preserving, unital processes are also called 
{\em bistochastic}. We identify bistochastic quantum operations, with non trivial Kraus representation, 
with {\em pure diffusion\/} (possibly with drift) while we associate dissipation with non-unital maps,
characterized  by a contraction parameter $\eta\neq 0$.

A typical example of bistochastic map is
a random sum of unitary operations
\begin{equation}
\su{S}(\rho)=\sum_\mu c_\mu U_\mu\rho U_\mu^\dag,
\end{equation}
with $U_\mu$ unitary and the trace preserving condition
\begin{equation}
\sum_\mu c_\mu=1,\ \ c_\mu\ge 0,
\end{equation}
also known as random unitary process (RUP). Diffusive noise in the
form of a RUP was studied in \cite{garma,garma2,nonn}, as a Gaussian sum of (normalized) translations in
phase space. When composed with a unitary map  
the whole noisy map 
can be interpreted as a coarse graining\cite{nonn,garma2} of the original map. 

Of course there can be other examples of bistochastic 
(or purely diffusive) channels which need not be RUP's.
The only necessary condition (which follows from \equa{eq:tp}) 
is that the Kraus operators be normal. 
In the context of quantum information noise is characterized by operations that can either take place on
single or multiple qubits, and consist in combinations of (tensor products of) Pauli matrices. While some
well known noises can be re-expressed as RUP's\cite{colo}, it is not the most general situation\cite{leung}.
\subsection{Sloppy Bakers}
Recently 
a model of an irreversible quantum baker map was presented by 
{\L}ozi\'nski et al.\cite{sloppy}. 
It consists of two steps: the usual quantization\cite{balazs,saraceno} of the baker
map which is unitary, and a projective measurement with a controlled 
translation in momentum. That is the bottom half of the torus 
is left untouched while the top half is translated an amount $\Delta/2$
The superoperator is non-unital except for the case where $\Delta=0$.
The invariant state is a uniform distribution on a reduced Hilbert space of $\text{\it
dim}=N(1-\Delta)$.
Thus the original invariant state of the unitary baker's map is reduced by a strip of area 
$\Delta$ (see Fig.~2 in \cite{sloppy}).

We do not reproduce the exact expression for the superoperator but the parameter $\eta$ for this map
can be shown to be
\begin{equation}
\eta=\Delta.
\end{equation}
As expected $\eta$ depends directly on the model's contraction parameter $\Delta$ in a simple way.

\begin{figure}[htb]
\includegraphics[width=6cm]{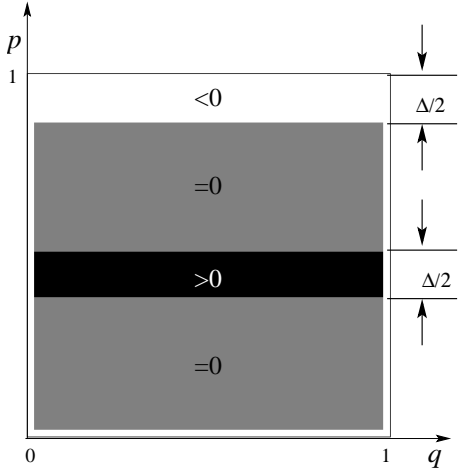}	
\caption{Sechematic representation of the Husimi function $\Gamma(z,z^*)$ for the noise used in
\cite{sloppy}. Dissipation takes place where $\Gamma(z,z^*)\neq 0$.\label{fig:slop}}
\end{figure}
The operator $A$ defined in \equa{eq:defA} for this model is
\begin{equation}
\Gamma=\frac{1}{N}\left(V^{-N\Delta/2}D_tV^{N\Delta/2}-D_t\right)
\end{equation}
where $V^{N \Delta/2}$ is a translation size $\Delta/2$ in momentum and $D_t$ is the projection onto
the upper half of the torus. For clarity in \fig{fig:slop}, instead of the Weyl representation of
$\Gamma$, the corresponding Husimi $\Gamma(z,z^*)$
is shown. The regions where $\Gamma(z)\neq 0$ are the regions where dissipation takes place.  
\subsection{Generalized amplitude damping noises}
	\label{sec:ditt}
A large variety of dissipative noise models which are derived from a microscopic Hamiltonian 
for a system (an oscillator \cite{liu}, a rotor \cite{dittrich}, a large spin \cite{haakebook,superradiance}, etc) in contact with a  bath in 
thermal equilibrium can be interpreted as generalized amplitude damping models and modeled by the following Kraus superoperator: 
\begin{equation}
\s(\rho)=\sum_k A_k\rho A_k^\dag
\end{equation} 
where
\begin{equation}
	\label{eq:aka}
A_k=\sum_{n}c_n^k\ket{n}\brac{n+k}
\end{equation}
is a combination of the transition operators 
\begin{equation}
	\label{eq:transition}
P_{ij}\stackrel{{\rm def}}{=}\ket{i}\brac{j}.
\end{equation}
Although the  basis states $ \ket{i}$ and the form of the coefficients  $ c_n^k$ 
depend on each particular problem, 
a common feature to all these models is that 
when $\s$ acts on the basis of the $P_{ij}$ 
the skewness $\alpha =(i-j)$ is conserved.
With the notation $P_{ij} = P_i^{\alpha} $ the evolution of the $P_l^{ \alpha}$  in each subspace labeled by $\alpha$
is given by 
\begin{equation}
\s P_i^{\alpha} = \sum_{l} m_{i,l}^{ \alpha}    P_{l}^{\alpha}
\label{subs2}
\end{equation}
with $m_{i,l}^{ \alpha} = c_l^{i-l} c_{l- \alpha}^{i-l *}$.

The preservation of the trace implies that $ \sum_{l=0}^{N-1} m_{i,l}^{ 0}= 1 $ while 
the unitality condition corresponds in addition to $ \sum_{i=0}^{N-1} m_{i,l}^{ 0}= 1 $. 
It is then clear that  in order to analyze the unitality of a noise of this type 
it is sufficient to study the properties of the $ (N \times N) $- matrix  $m_{i,l}^{ 0}$, unitality 
corresponding to the bistochasticity of this matrix.

It is immediate to see that if the matrix  $m_{i,l}^{ 0}$ is either symmetric, or symmetric with respect 
to the diagonal $ i+j= N-1 $ stochasticity implies bistochasticity. 
 
A symmetric matrix corresponds 
to having a purely diffusive reservoir, i.e., a thermal bath at infinite temperature, exchanging quanta 
with the system  in both directions at equal rate. 

In the second case the matrix elements $m_{i,l}^{ 0}$  only depend on $( i-j ) $,  
implying that in \equa{eq:aka} the coefficients   $c_{i}^{k}= c(k)$. Expanding the transition 
operators in the basis of the translations on the torus $T_{(q,p)}$ (as defined by
Schwinger\cite{Schwinger}),
it is easy to see that the superoperator $\s$ is diagonal in this basis
\begin{equation}
\s \rho= \frac{1}{N^2}\sum_k c(-k) T_{(k,0)}\rho T^\dag_{(k,0)}
\end{equation}
that is, it can be written as an incoherent sum of translations in phase space.

If we compute $\eta$ for this type of models we get:
\begin{equation}
\eta =  \sum_{l} (\sum_{i} m_{i,l})^2 -1.
\end{equation}

For the dissipative map studied by Dittrich and Graham in \cite{dittrich} (adapted to a finite dimensional Hilbert space), 
this gives in the limit of zero temperature (to second order in $\gamma$)
\begin{equation}
	\label{eq:etaDit}
 \eta = \gamma^2(N-1),
\end{equation} 
 where $ \gamma $ is the friction parameter 
(differential rate of loss of action) of the corresponding classical map (see Eq.~(2.4) in \cite{dittrich}).
\section{Simple Dissipation Channel}
	\label{sec:SDC}
We propose a family of non-unital noise channels to serve as simplified models of dissipative
processes like the one described in Sec.~\ref{sec:ditt}
Let 
\begin{equation}
	\label{eq:s-model}
\su{D}_\epsilon(\rho)=(1-\epsilon)\rho+\epsilon\sum_{ij}p_{ij}P_{ij}\rho P_{ij}^\dag,
\end{equation}
with $P_{ij}$ as in \equa{eq:transition} and $p_{ij}$ real and positive. Equation (\ref{eq:s-model}) is in
Kraus form and thus $\su{D}_\epsilon$ is CP. The TP condition (\ref{eq:tp}) is
\begin{eqnarray}
\sum_{ij}p^T_{ij}P_{ij}^\dag P_{ij}&=&\II\nonumber\\
\sum_iP_{ii}\sum_jp^T_{ij}&=&\II
\end{eqnarray}
\begin{figure}[!h]
\begin{center}
\includegraphics[width=6cm]{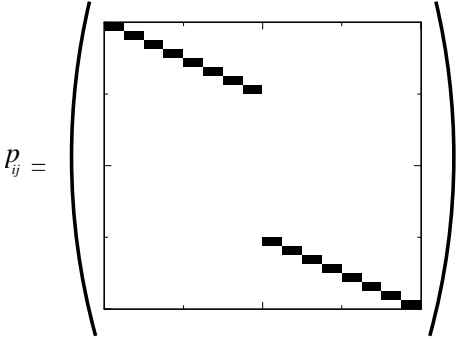}		
\caption{Graphical  representation of the structure of the matrix $[p_{ij}]$ with $N=32$ and $\alpha=0.5$. The color scheme is as follows: white means
null value and a black square equals one element of value equal to 1. 
\label{fig:pij}}
\end{center}
\end{figure}
\begin{figure}[htb!]
\begin{center}
\includegraphics*[width=8cm]{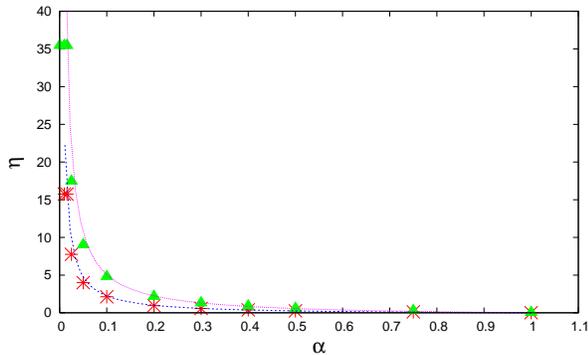}	
\caption{(Color online) The contracting parameter $\eta$ as a function of $\alpha$ for the dissipation model of
\equa{eq:alpha} for two different values of $\epsilon$: ($*$) $\epsilon=0.5$, ($\blacktriangle$)
$\epsilon=0.75$.\label{fig:eta}}
\end{center}
\end{figure}
\begin{figure*}[htb!]
\begin{center}
\includegraphics*[width=16cm]{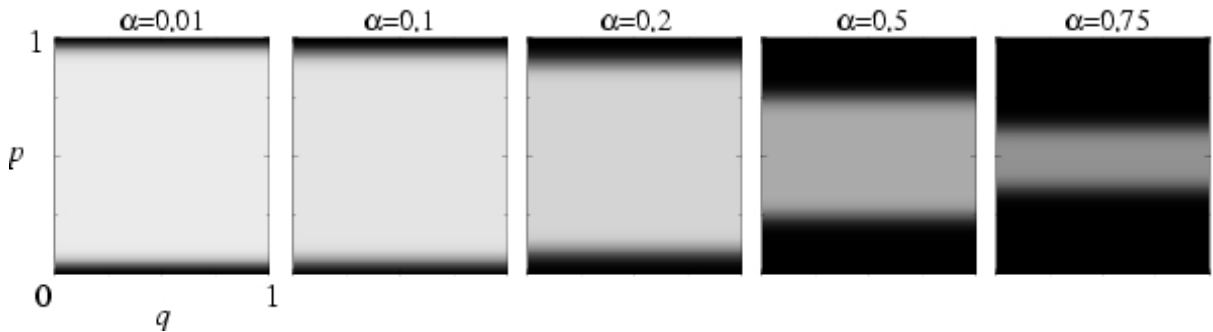}  
\caption{Plot of the Husimi function $\Gamma(z,z^*)$ for $N=64$ (in gray scale, black is
maximum and white minimum)
and different values of $\alpha$.
Lighter regions indicate local contractive regions.\label{fig:compresion}}
\end{center}
\end{figure*}
where $T$ means `transpose'.Therefore the matrix of
coefficients $p_{ij}$ should be stochastic. If it is doubly stochastic then the noise is unital.
This is a model which is diagonal in the $P_{ij}$ representation and decoherence
appears as a reduction by a factor $(1-\epsilon)$ of the non-diagonal terms of the density matrix.

We can {\em create\/} dissipative models just by using non-symmetric stochastic
matrix of coefficients $p_{ij}$.
Consider the family
\begin{equation}
	\label{eq:alpha}
\su{D}_{\epsilon\alpha}(\rho)=(1-\epsilon)\rho+\epsilon\sum_{i,j=-N/2}^{N/2}P_{[\alpha i]\, i}\rho P_{[\alpha i]\,i}^\dag,
\end{equation}
with $\alpha\in[0,1)$ and the index $[\alpha i]$ is the integer part of $\alpha i$, 
as a function of the parameter $\alpha$ and the negative values are taken mod$(N)$. 
In \fig{fig:pij} there
is an illustration of the structure of the matrix $[p_{ij}]$ of \equa{eq:s-model} for the case $\alpha=0.5,
N=32$. 

For all values of $\alpha$, the first term attenuates uniformly the off-diagonal 
elements of the density matrix, introducing decoherence.
The second term which acts on the probabilities 
(diagonal elements) takes the system to lower states. The permitted transitions of the
diagonal elements are determined by the parameter $\alpha$ but its dependence on the parameter 
$\epsilon$, which accounts for the coupling strength of the system to the environment, 
becomes important when the contraction is considered.

If $P{ij}$ are transitions in the computational basis of a
state of say $k=\log_2 N$ qubits, then the noise can be interpreted as follows. With
probability $(1-\epsilon)$ it leaves the initial state untouched while with probability
$\epsilon$ it induces errors in the form of transitions which depend on the parameter
$\alpha$.
For $\alpha =1$ the superoperator is unital and corresponds to a phase 
damping channel for $k=\log_2 N$ qubits . If  $\alpha <1 $  this very simple model 
captures some important features of the amplitude damping noises (an example of which is
described in \sect{sec:ditt}).
On the other hand, if we take $P_{ij}$ to be
transitions in \textit{momentum\/} and since the invariant state (for all $\alpha\neq 0$) is
$\rho^*=P_{00}=\ket{p=0}\brac{p=0}$ then the noise has the effect of a friction. 
As is shown later, in phase space representation the parameter $\alpha$ 
is related to the region over which dissipation acts. 

The simple form of \equa{eq:alpha} allows to compute the contraction parameter easily as
\begin{eqnarray}
\eta&=&N\tr\left[(\su{D}_{\epsilon\alpha}(\II/N)-\II/N)^2\right]\nonumber\\
    &=&\frac{\epsilon^2}{N}\tr\left[\II-2\sum_iP_{[\alpha i],[\alpha i]}+
    \sum_{ij}P_{[\alpha i],[\alpha i]}P_{[\alpha j],[\alpha j]}\right]\nonumber\\
    &=&\frac{\epsilon^2}{N}\tr\left[\sum_{ij}|\bra{\alpha j}\ket{\alpha i}|^2-N\right]
\end{eqnarray}
where $\sum_iP_{[\alpha i],[\alpha i]}=N$ and for simplicity we drop the $[\ ]$ symbol inside
the kets and bras. Now $\bra{\alpha j}\ket{\alpha i}=0,1$ so the square can be dropped and in
order to calculate the sum we approximate by continuous variables as
\begin{eqnarray}
	\label{eq:eta-analytic}
\sum_{ij}\bra{\alpha j}\ket{\alpha i}&\sim& \iint\delta(\alpha(x_i-x_j))\simeq 
{1\over\alpha}\iint\delta(x_i-x_j)\nonumber\\
&\sim&{1\over\alpha}\sum_{ij}\delta_{ij}=\frac{N}{\alpha}.
\end{eqnarray}
Therefore we get
\begin{equation}
\eta=\epsilon^2\left(\frac{1-\alpha}{\alpha}\right).
\end{equation} 

In figure \fig{fig:eta}we plot $\eta(\alpha,\epsilon)$ for two different values of $\epsilon$. We see
that the computed points fit exactly the analytic expression. Moreover the saturation value,
\begin{equation}
\eta(\alpha=1/N,\epsilon)=\epsilon^2(N-1).
\end{equation}
for small $\alpha$ can be understood as follows. Since the dimension of Hilbert space is $N$ then
the integer part $[\alpha\,i]$ ($i=0,\ldots N-1$) for any value of $\alpha<1/N$ is zero. 
We notice that the dependence on the coupling parameter $\epsilon $ is 
the same as the one given in \equa{eq:etaDit} for the dissipative map of \cite{dittrich}, if 
one identifies it with the friction parameter $\gamma$. In addition, it can be easily seen by taking the mean values 
of $p$ and $q$ that this operation implies the following dissipative map
in the classical limit
\begin{equation}
	\label{eq:cldiss}
\begin{array}{rcl}	
\bar{q}'&=&\bar{q}\\
\bar{p}'&=&(1-\epsilon)\bar{p}
\end{array}
\end{equation}
where the bar indicates mean values.

For this model the operator $\Gamma$ defined in \equa{eq:defA} is
\begin{equation}
	\label{eq:gamma}
\Gamma=\epsilon\sum_{i=-N/2}^{N/2} \left(P_{[\alpha i][\alpha i]}-P_{ii}\right).
\end{equation}
So the region of dissipation (determined by a negative value of $A$) has 
area equal to $1-\alpha$, and occupies the central region of the unit square. 
In figure \fig{fig:compresion} the Husimi function of the operator $\Gamma$ 
is represented taking different values of $\alpha$ (and $\epsilon=0.5$). The light region represents the
area where contraction takes place and corresponds to a negative value of $\Gamma$. For $\alpha\sim 0$
the contraction is uniform over (almost) all phase space, except over the state $\ket{0}$ of momentum.
\section{Composition with a Unitary Process}
	\label{sec:unitary}
 Following recent works\cite{bian,garma, garma2,nonn,sloppy,braun,braunbook} we study the effect of the dissipative noise channel described in \sect{sec:SDC} 
when composed with a unitary
map. 
As an example we take the
quantum version of the standard map on the torus. We suppose that to a good approximation the whole noisy propagation takes
place in two steps
\begin{equation}
\su{S}(\rho)=\su{D}_{\epsilon\alpha}(\su{U}(\rho))
\end{equation}
where $\su{U}(\rho)=U\rho U^\dag$ is the unitary step. The two-step scheme can can be used, in the master equation if the
Hamiltonian part commutes, to some desired order in $\hbar$ with the non-Hamiltonian part. In the quantum operation
formalism these scheme is suitable in the case for example where unitary evolution 
takes place in so short times that the noise is negligible (e.g. the micro-maser, a billiard where the interaction with the
walls is very short and the evolution inside is dissipative).
\begin{figure}[htb]

\includegraphics[width=8.25cm]{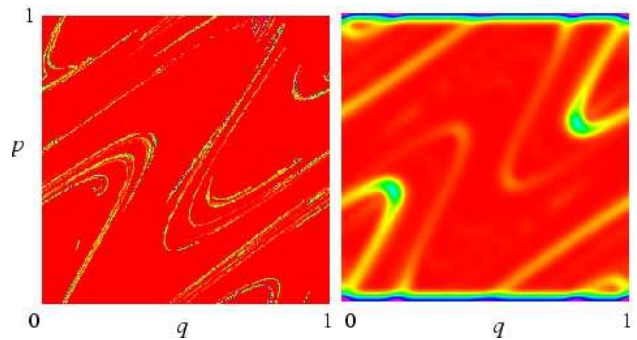}	
\caption{(Color online) Left: histogram representation of the 
attractor for the classical standard map (\equa{eq:std-map} with $k=0.065,\delta=0.6$) 
with dissipation. Right: Husimi representation for invariant state for the
quantum standard map with the dissipative noise described in \sect{sec:SDC} ($\alpha\sim 0,
\epsilon=0.4$\label{fig:comp}}
\end{figure}
The unitary map chosen is the quantum version of the standard map on the torus 
\begin{eqnarray}
	\label{eq:std-map}
q'&=&q+p'\\
p'&=&\delta p-2 \pi k \sin(2 \pi q)
\end{eqnarray}
where the factor $\delta<1$ on the momentum acts as friction. 

\begin{figure*}[!!htb]
\begin{center}
\includegraphics[width=17.cm]{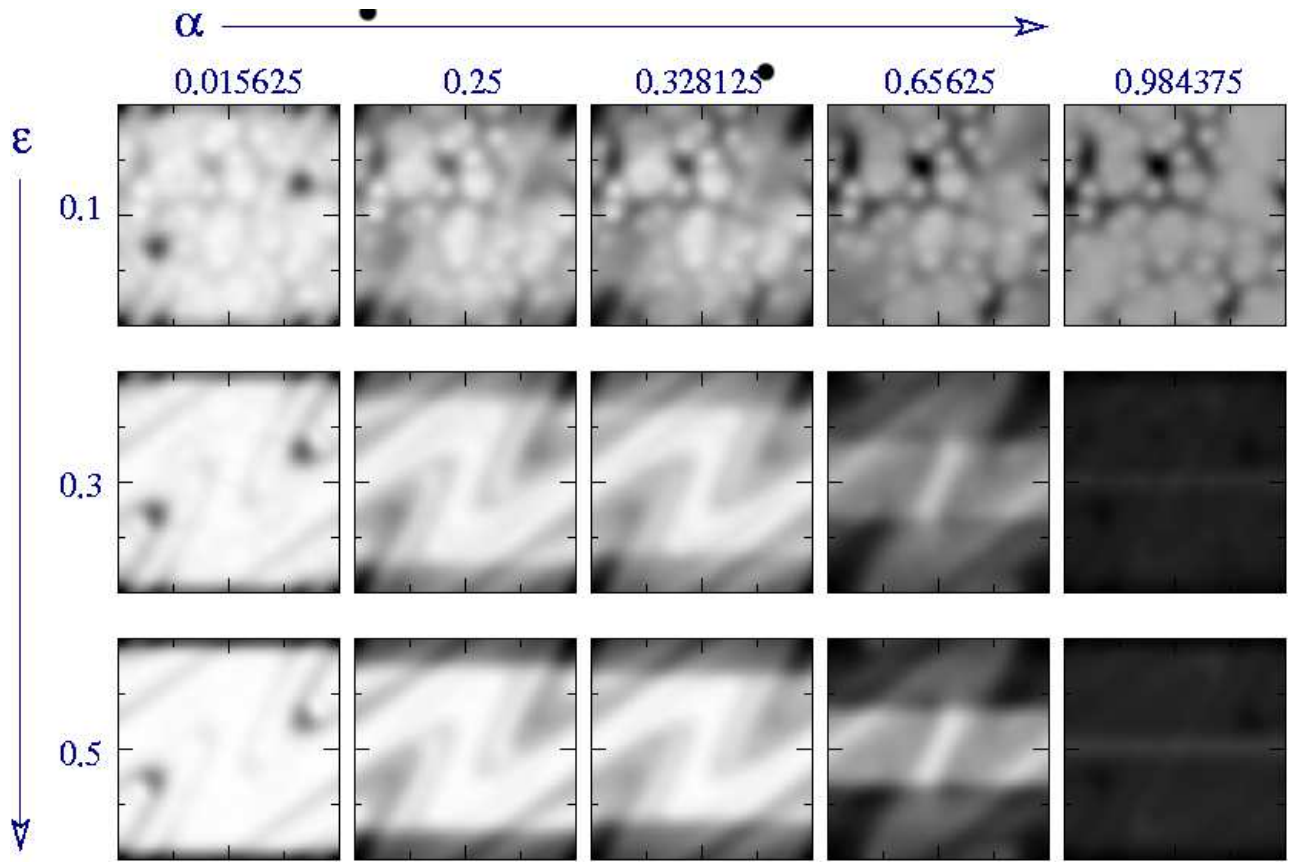}		
\caption{Husimi representation (same color scheme as \fig{fig:compresion}) of the 
invariant state for the standard map with dissipation for different values of $\alpha$ and $\epsilon$ and
$N=64$. The difference between $\alpha\approx 0$ (dissipation) and $\alpha\approx 1$ (diffusion) can be observed.
(Axes and units are the same as in FIGS.~\ref{fig:compresion} and \fig{fig:comp}.)
\label{fig:eps-alpha}}
\end{center}
\end{figure*}
In \fig{fig:comp} the classical and quantum invariant states are plotted. 
The classical attractor on the left is represented as density in phase space. 
On the right is the Husimi distribution of the invariant state for the quantum version of the map followed by the noise of
\equa{eq:alpha} for $\alpha\sim 1/N$ ($N=64$).
Since the dissipation terms in 
Eqs.~(\ref{eq:cldiss}) and (\ref{eq:std-map}) are the same when $\delta=(1-\epsilon)$, as expected
the quantum invariant state exhibits the structure of the classical attractor.

The shadow on the lower and upper edge of the torus can be explained from the definition of $\Gamma$, \equa{eq:gamma}, 
for this noise (corresponding to the rightmost image in \fig{fig:compresion}).
The same argument can be used to explain \fig{fig:eps-alpha}. As $\alpha$ grows the region of dissipation becomes smaller as well as the
contracting parameter $\eta$. The noise becomes increasingly similar to a unital operation, at least on the upper and lower bands of
width $\alpha/2$. So over the black shaded regions of \fig{fig:eps-alpha} the noise acts like a generalized phase damping 
channel \cite{colo}, 
where the preferred basis are the momentum projectors $P_{ii}$ with $i=\{0,\ldots,\alpha N/2\}\cup\{(1-\alpha/2)N-1,\cdots,N-1\}$,
while on the lighter region, a dissipation of the type of \equa{eq:cldiss} acts and 
the attractor is uncovered.   
\section{Conclusions}
The fundamental difference between unital and non-unital processes was explored. 
In analogy to the classical limit of the master equation, which relates the commutator of the Lindblad operators to the vector drift in
its Fokker-Planck limit, we
defined a parameter that measures non-unitality and characterizes dissipative quantum operations. 
As an example we proposed a noise channel that
displays in a simple  way the essential features of decoherent and dissipative processes. 

The non-unitality of the superoperator is related to a traceless Hermitian operator whose phase space distribution
gives a local image of the dissipation process. 
This operator is independent on the Kraus representation and can give a useful
insight into the dissipative properties of the superoperator 
when a phase space description or a semiclassical limit is not available, like for general noise
channels in quantum information.

\acknowledgments
Partial support for this work was provided by ANCyPT and CONICET. I.G.-M. and M. S. 
thank the hospitality at the 
{\em Center for Nonlinear and Complex Systems\/} (Como) and G. Benenti for fruitfull
discussions.

\end{document}